\documentstyle[twocolumn,prl,aps]{revtex}
\begin{document}
\draft
 
\title{Critical Behaviour of Superfluid $^4$He in Aerogel }
 
\author{K. Moon} 

\address{
Department of Physics, University of California,
Davis, California 95616
}

\author{S. M. Girvin}

\address{
Department of Physics, Indiana University,
Bloomington, Indiana 47405
}

\date{\today}
\maketitle
 
{\tightenlines
\begin{abstract}
We report on Monte Carlo studies of the critical behaviour of superfluid
$^4$He in the presence of quenched disorder with long-range
fractal correlations. According to the
heuristic argument by Harris, uncorrelated disorder is 
irrelevant when the specific heat critical exponent $\alpha$ is negative,
which is the case for the pure $^4$He.
However, experiments on helium in aerogel
 have shown that the superfluid density critical 
exponent $\zeta$ changes. We hypothesize that 
this is a cross-over effect due to the fractal nature of aerogel. 
Modelling the aerogel as an 
incipient percolating cluster in 3D 
and weakening the bonds at the fractal sites, we perform XY-model simulations,
which demonstrate an increase in $\zeta$ from $0.67 \pm 0.005 $ 
 for the pure case to an apparent value of $0.722\pm 0.005$ 
in the presence of the fractal disorder, provided that the
helium correlation length does not exceed the fractal correlation length.
\end{abstract}
}
 
 
\narrowtext
It is generally believed that the superfluid
transition ($\lambda$-point) of pure $^4$He belongs to the 
classical 3D-XY model universality
class. Near the critical temperature $T_c$, the superfluid density scales
as $|T-T_c|^\zeta$, where $\zeta$ is measured\cite{ahlers_review}
 to be $0.674\pm 0.003$.
There has been considerable interest in studying the superfluid transition 
for helium in a variety of porous media.\cite{reppy_review}
  One such system is Vycor which
is a random glass with a porosity of order 30\%.  Remarkably, the critical
exponent $\zeta$ is found to be unchanged by the seemingly huge perturbation
represented by the glass.  This however is the result expected according to
heuristic arguments of Harris\cite{harris} and more rigorous recent
work\cite{chayes-chayes-fisher} which shows that weak, uncorrelated
randomness is irrelevant at the unperturbed
critical point provided that the specific heat exponent is negative.
The exponent\cite{ahlers_review}
$\alpha \approx -0.026\pm 0.004$ is indeed negative for $^4$He, however 
Narayan and Fisher\cite{Narayan} have argued that since $\alpha$ is only
slightly negative, the crossover to the pure 3D XY critical regime is almost 
logarithmically slow.  Until recently Vycor was the only porous medium 
in which $\zeta$ was observed to be unchanged.  New experiments however 
appear to have added a second material, porous gold (which also has
short-range correlations), to the list.\cite{chan_priv_comm}

Remarkable results have been obtained for the aerogel system.  
Aerogel is
a fractal silica `dust' with porosities of 95-98\% or higher.  Despite the
fact that aerogel is almost entirely empty space, and that it (unlike Vycor)
has only a tiny effect on the critical temperature, the exponent $\zeta$ is
apparently quite distinctly
shifted\cite{reppy_zeta,ahlers_zeta,chan_priv_comm}
 to approximately 0.75 (or larger), 
indicating that the nominally weak aerogel 
perturbation is  relevant at the 3D XY critical point
and aerogel (perhaps) produces a new universality class.  The specific heat
data suggests that either hyperscaling is violated or the amplitude
ratio is exceptionally small.\cite{reppy_sp_heat,ahlers_sp_heat}
In addition to shifting
$\zeta$ and $\alpha$, aerogel dramatically
changes the topology of the $^3$He-$^4$He mixture phase 
diagram.\cite{topology_expt,numerics}  

It is known that long-range correlations can make disorder relevant, even
when $\alpha$ is negative.\cite{Weinrib,grinstein}  Weinrib and
Halperin\cite{Weinrib} have demonstrated this
for the special case of weak, {\it gaussian
distributed\/} disorder with long-range correlations. 
Li and
Teitel\cite{Teitel} have looked at a model of {\it non-weak\/} 
(broadly distributed) but
{\it uncorrelated\/} 
disorder and find an apparent increase in the exponent $\zeta$
to a value which depends on the exponent in the algebraic
divergence of the disorder strength distribution at weak disorder.

Machta\cite{Machta}
 has considered a model of aerogel as a relatively uniform medium
filled with pores on many different length scales.  This model was
motivated by early experiments which saw two transitions, one at the
usual bulk temperature and one at a slightly lower temperature.  Recent
improvements in aerogel synthesis techniques appear however to have eliminated
the larger pores and inhomogeneities so that now only a single transition 
is observed at a temperature slightly below the bulk 
value.\cite{chan_priv_comm}  Recently,
Huang and Meng have examined a mean-field
theory in a percolating cluster system.\cite{Huang}

In order to 
investigate this fascinating subject, we have performed extensive Monte
Carlo simulations on the 3D XY model for three cases: i) no disorder,
ii) uncorrelated disorder, and iii) fractal disorder.  We consider lattice
sizes up to $24^3$, and use the Wolff algorithm to minimize
the otherwise severe effects of critical slowing down.\cite{Wolff,Janke,Ilpo}.
The model is defined by compact phase variables $\{\theta\}$ on sites
of a simple cubic lattice
\begin{equation}
S[\theta]=-\sum_{\bf r,\delta}\frac {K_{\bf r,\delta}}{T}
\cos\left[\theta({\bf r})
-\theta({\bf r+\delta})\right].
\end{equation}
We measured the disorder-averaged
superfluid density $\rho_s$ as a function of temperature 
and the system size using the usual Kubo formula expression.\cite{min-chulPRB}
Defining one `sweep' as growing and reorienting a cluster of spins 
(on the order the system size when near the critical point)
with the Wolff algorithm, typical runs involved $5\times10^4$ warm-up sweeps,
and $2\times10^5$ production sweeps, with measurements taken every 200 sweeps.
Results were averaged over typically 100 disorder realizations.  

A finite-size scaling analysis is crucial to the accurate determination 
of the critical temperature and exponents.  The scaling ansatz assumes
that $\rho_s$ has the form\cite{min-chulPRB,Teitel}
\begin{equation}
\rho_s={T}{L^{-\Omega}} G(L/\xi)=
{T}{L^{-\Omega}} {\tilde G}[(T-T_c) L^{1/\nu}],
\end{equation}
where $\Omega =  (2 - \alpha)/\nu -2$.
Hence, the dimensionless combination $\rho_s L^\Omega/T$ 
is scale invariant at the critical point $T_c$.   If we assume
that hyperscaling holds, then we have $\Omega = d-2 = 1$, otherwise
$\Omega$ is {\em a priori} unknown.
In the inset of Fig.(\ref{fig1}) we determine $T_{\rm c}=2.156\pm 0.001$ by
plotting $\rho_s L/T$    
vs. temperature for uncorrelated disorder    
\begin{equation}
K_{\bf r,\delta}=1+\delta K_{\bf r,\delta} \; ;  
\; \delta K_{\bf r,\delta} \in [-\Delta,\Delta] \; \; {\rm   and  }\; \;
\Delta<1 .  
\end{equation}
of relatively large (but bounded) strength $\Delta=0.7$.
Similar calculations for the pure (disorder-free)
case yield distinctly larger value $T_{\rm c}\cong 2.203\pm 0.001$. 
In the main part of 
Fig.\ref{fig1}, ${\tilde G}$ is plotted as a function of the scaling 
variable
$(T-T_c) L^{1/\nu}$ with the value
$\nu=0.667\pm 0.005$ which gives the best data collapse
onto a single universal scaling curve. Clearly, this unchanged exponent
is consistent with the Harris criterion
and the experimental observation in Vycor that uncorrelated randomness does
not change the universality class.\cite{comment_log} 
In order to cross check the above result, we have also measured the
magnetization $m$ and computed the Binder
ratio,\cite{Binder,Huse}
which is automatically scale invariant at the critical point,
\begin{equation}
U_4=1-\biggl[\frac{<m^4>}{3<m^2>^2}\biggr]_{\rm ave}.
\end{equation}
This allows one to estimate the critical  
exponent $\nu$ without assuming a value for $\Omega$ as was necessary
in the case of the superfluid density.  Plotting $U_4$ against
the variable $(T-T_c) L^{1/\nu}$, we found good scaling
 for the same values of $T_c$ and $\nu$ obtained from the scaling
of $\rho_s$.

We turn now to a discussion of fractal disorder.  A material with
fractal dimension $D_{\rm f}$ has a mass that scales with length
like $M\sim L^{D_{\rm f}}$.  If $D_{\rm f}<3$  then (real) objects
can never be fractal beyond some finite correlation length $\xi$ because
otherwise the density would vanish.
Static structure factor
measurements\cite{fractal_struct} indicate that acid-catalyzed aerogel has 
fractal dimension $D_{\rm f} \sim 2.4 -2.5$ over a wide range of
length scales from $\sim 6\AA$ out to roughly $\xi \sim 600\AA$.  
Base-catalyzed aerogel (typically used in the helium experiments) is believed 
to have a somewhat lower fractal dimension and a lower range of length
scales.  Various measurements of the `fracton' 
vibrational properties of aerogels however place a {\it lower bound\/} on
the correlation length for the {\em connectivity} that is {\em at least}
an order of magnitude larger.\cite{connectivity}
  The connectivity structure may be important
because the closed vortex loops in the helium are presumably attracted
to the aerogel strands and are thus sensitive to the connectivity.
One of the central mysteries shown up
by the experiments is the following.
In the critical regime with reduced temperature
$t \sim 10^{-5}$ the correlation length of the helium is expected to
be larger ($\sim 1 \mu{\rm m}$) than even the (estimated) connectivity
length scale.  Nevertheless no evidence of a cross-over
to the uncorrelated disorder regime is evident in the full-pore
experiments.\cite{reppy_review}
Very recently however, Crowell et. al.\cite{reppy_preprint}
 have performed experiments in
the regime of lower helium densities where even larger correlation
lengths can be obtained.  They find evidence of a possible cross-over
to the uncorrelated disorder regime with a lower value of $\zeta$.

We have considered the possibility that the deformability of the tenuous
aerogel structure is relevant.  On length scales beyond the fracton
correlation length, aerogel acts to sound waves
like a relatively homogeneous system with a low speed of sound ($\sim
100$m/s) despite its very low mass density, indicating that its
compressibility is nearly $10^6$ times that of ordinary glass. 
The aerogel is known to be sufficiently flexible that it modifies
the collective sound mode dispersion.\cite{maynard,ahlers_zeta}
On scales
beyond the fracton correlation length, it is reasonable to argue that the
aerogel density fluctuations $\delta\Phi$ act as, simple, uncorrelated
local {\em annealed} disorder coupling to the magnitude of the helium order
parameter in a Ginsburg-Landau theory with action
\begin{eqnarray}
S &=& \frac{1}{2\kappa} (\delta\Phi)^2 
+ [\frac{1}{2} - h\, \delta\Phi] |\nabla\psi|^2\nonumber\\
&+& [\alpha + g\, \delta\Phi]|\psi|^2 
+ [\lambda +k\, \delta\Phi]|\psi|^4.
\end{eqnarray}
Integrating out $\delta\Phi$ for small $g,h,k$ produces only irrelevant
couplings.  For stronger couplings however, the system can,
in the right circumstances, be driven to a tricritical point, beyond
which the transition is first order.  This is precisely
what happens in $^3$He-$^4$He mixtures where it is a good approximation to
treat the $^3$He impurities as annealed disorder.\cite{reppy_preprint,ahlers_review}  
This confirms the idea that deformability of the aerogel should be
irrelevant. It should be noted however that it is probably more
physically correct to include the constraint $\int d^3r \delta\Phi = 0$ which
would lead to the slow logarithmic case of Fisher 
renormalization of the critical exponents.\cite{fisher_renorm}
The possibility that this may account for the peculiar features 
of the specific heat data should be looked into in more detail.  

We seem to be
left only with the possibility of aerogel as quenched disorder whose
fractal character extends beyond the lower bound set by the fracton cutoff.
A variety of schemes have been used to  model the aerogel 
structure.\cite{Machta,Huang,numerics,stauffer}
We have chosen a simple percolation model\cite{stauffer} for the fractal 
structure (probably more appropriate for acid-catalyzed than base-catalyzed
aerogel).  We generate a critical
percolation backbone on the 3D lattice by randomly occupying lattice sites
with probability $p=p_{\rm c}$ and keep only the largest connected 
cluster.
In addition, we selected only fractal realizations with porosity
in a narrow window centered on the median value 
in order to reduce the sample-to-sample
fluctuations in the disorder strength.  
We confirmed that these objects had the known fractal dimension\cite{stauffer}
 $D_{\rm f} \sim 2.5$.
 
In order to be able to do finite size scaling, while avoiding the scale
dependence of the porosity, 
a single large cluster was generated on an $L_0=48$ lattice
(giving a porosity of about 95\%)
and divided into smaller subsystems of size $L=8,12,16,24$.
Simulations were performed for different subsystems with periodic
boundary conditions and averaged over the
subsystems and different fractal realizations.
The bond strength $K$ on the fractal was arbitrarily reduced from unity
 to 0.26.
The inset of Fig.(\ref{fig2}) shows $\rho_s L/T$ vs. $T$, and appears
to give a clear fixed point with $T_c$ estimated to be 
$2.1834\pm 0.001$ 
which is closer to the 
pure critical point than for the uncorrelated disorder model,
 since the porosity of the fractal is so high.
In the main part of Fig.(\ref{fig2}), we plot
 ${\tilde G}$ vs. the scaling variable
and find that the (apparent) critical exponent $\zeta$ increases to 
$0.722\pm 0.005$.  We have confirmed this result with measurements
of the Binder ratio.  Unlike the case of uncorrelated disorder,
we observed a slow drift downward of $U_4$ with system
size at the previously determined $T_c$.  Taking this out by 
scaling the data by the factor $U_4(T_c,0)/U_4(T_c,1/L)$ yields essentially
perfect data collapse with $\nu=0.72\pm 0.007$,   
as shown in Fig.(\ref{fig3}).
Assuming a violation of hyperscaling gives
an anomalous dimension to the superfluid density
$\rho_s \propto L^{-(1+\theta)} {\tilde G}[(T-T_c) L^{1/\nu}]$,
we can place an approximate upper bound $|\theta| \le 0.06$. 

It is enlightening to compare the present results to a model with
disorder of lower dimension, namely infinitely long
columnar defects.\cite{cha_columnar}
Recent work on this model
indicates that $\nu_\perp=\zeta_\perp\sim 1$ is even larger than
for the present model. 
This must be the case in order to
 satisfy the rigorous lower bound\cite{chayes-chayes-fisher}, 
since this model {\it does\/} truly represent a new universality class,
and not simply a cross-over.
The superfluid density measured parallel to the columns has an
even larger exponent ($z \equiv \nu_\parallel/\nu_\perp \sim 1.07$).

In conclusion, we have argued that the apparent increase in the superfluid
density exponent in aerogel can not be due to a true change of universality
class but must be a cross over effect in the regime where the helium
correlation length is less than the (apparently large, but
{\it necessarily finite\/}) correlation
length for the disorder.  We have performed Monte Carlo simulations in
this regime for a percolation cluster model of
fractal disorder and find an increase in the effective exponent to 
$\zeta = 0.722\pm 0.005$   
which appears to be roughly consistent with experiment.  However,
we see no
apparent violation of hyperscaling,  and attempts to confirm the unusual
behavior of the experimental specific heat in our model have proved
too difficult computationally at this time.

\acknowledgements{
It is a pleasure to thank G. Ahlers,
T. Witten, D.S. Fisher, D. Huse, M.W. Chan, 
C. Hanna, L. Kadanoff, N. Read, J. Machta, M. Cha, G. Zimanyi, R. Scalettar,
R. Singh, S. Teitel, J. Reppy, and S. Renn for helpful discussions.  
The work at Indiana was supported by the NSF through 
grant No.  DMR-9416906. 
The work at UC-Davis was supported by NSF Grant No. DMR 92-06023 and by the
Los Alamos National Laboratory through a LACOR grant.
}

\begin{figure}
\caption{
Universal scaling function ${\tilde G}$ vs.
the scaling variable $(T-T_c) L^{1/\nu}$ for the case of uncorrelated
disorder. The critical exponent $\nu$ is estimated to be $0.667\pm 0.005$.    
The inset shows 
the dimensionless superfluid density $\rho_s L/T$ plotted against
temperature.
}
\label{fig1}
\end{figure}

\begin{figure}
\caption{Universal scaling function ${\tilde G}$ for the case of
95\% porosity fractal disorder.  The critical exponent $\nu$ is
estimated to be $0.722\pm 0.005$.  The inset shows 
the dimensionless superfluid density $\rho_s L/T$ plotted against
temperature.
}
\label{fig2}
\end{figure}

\begin{figure}
\caption{Universal scaling curve for the (size-corrected) Binder ratio
for the case of fractal disorder.
The choice of the critical exponent $\nu=0.72\pm 0.007$  
 makes the data for different
system sizes collapse onto a single curve and agrees with the conclusion
drawn from analysis of the superfluid density.
} 
\label{fig3}
\end{figure}


\begin{references}
\bibitem{comment_log}  It is possible that there is a 
 logarithmic correction to scaling\cite{Narayan} 
 visible in our data, but it is very weak.
\end{references}
\end{document}